\documentclass[aps,prd,10pt,nofootinbib,twocolumn]{revtex4-1}
\usepackage{amsmath,amssymb,amsfonts,dsfont,mathrsfs,amsthm,mathtools}
\usepackage{graphicx}
\usepackage{hyperref}
\usepackage{siunitx}
\usepackage[english]{babel}
\hypersetup{linktocpage,colorlinks=true,urlcolor=blue,linkcolor=blue,citecolor=blue}

\begin{document}

\title{Symmetries and conserved quantities with arbitrary torsion: A generalization of Killing's theorem}

\author{Christian Peterson}

\author{Yuri Bonder}
\email{bonder@nucleares.unam.mx}

\affiliation{Instituto de Ciencias Nucleares, Universidad Nacional Aut\'onoma 
de M\'exico\\ Apartado Postal 70-543, Cd.~Mx., 04510, M\'exico}

\begin{abstract}
When spacetime torsion is present, geodesics and autoparallels generically do not coincide. In this work, the well-known method that uses Killing vectors to solve the geodesic equations is generalized for autoparallels. The main definition is that of T-Killing vectors: vector fields such that, when their index is lowered with the metric, have vanishing symmetric derivative when acted with a torsionfull and metric-compatible derivative. The main property of T-Killing vectors is that their contraction with the autoparallels' tangents are constant along these curves. As an example, in a static and spherically symmetric situation, the autoparallel equations are reduced to an effective one-dimensional problem. Other interesting properties and extensions of T-Killing vectors are discussed.
\end{abstract}

\maketitle

\section{Motivation}

General relativity (GR) is accepted as the theory of gravity. It is an appealing theory from the mathematical and philosophical points of view and many of its predictions have been empirically verified \cite{Will}. It states that the spacetime metric is dynamical and, from energy-momentum conservation, I can be shown that free point-like test particles follow geodesics \cite{Papapetrou}. In addition, GR assumes that the connection is torsionless, i.e., that the derivative operator associated with such a connection vanishes when acting twice, anti-symmetrically, on any scalar field. Geometrically, this means that infinitesimal geodesic parallelograms close \cite{Nakahara}.

Nevertheless, there are several arguments that suggests that one should look for a gravity theory beyond GR. First, within GR, there are singularities in physically meaningful solutions \cite{Singularities}. Second, for GR work at cosmological scales, one needs to supplement the Standard Model with dark matter \cite{DM}, a mechanism to produce inflation \cite{Inflation}, and there are issues with dark energy even if it is assumed to a cosmological constant \cite{Weinberg:1988cp}. And third, the best description of matter, which is a source of gravity, is quantum mechanics. Therefore, one should be able to predict the behavior of spacetime when sourced by matter in highly nonclassical states; this is the problem of quantazing gravity, for which there are several candidates \cite{Oriti}.

To tackle some of these issues, people have considered modifications to GR within the geometrical paradigm; the resulting theories are known as modified gravity theories. The most popular kind of modified gravity theories are the so-called $f(R)$ theories \cite{fofR} where the dynamical variables are those of GR but the action is extended to be a function of the curvature scalar.

On the other hand, one of the earliest studies on modified gravity theories were done in the context of Riemann--Cartan geometries, where the connection is torsionful, and which were rediscovered when tryin to gauge the Poincar\'e group \cite{Kibble,Sciama}. There are several reviews and books on this topic, for example \cite{Hehl:1976,Blagojevic:2002du,Obukhov:2006gea,Blagojevic:2013xpa}. The so-called Einstein--Cartan theory is the simplest theory with nontrivial torsion. The action of such a theory has the same functional form than that of GR and, in this theory, torsion does not propagate. Indeed, all predictions coincide with those of GR outside spin-polarized matter. However, inside this type of matter, there are deviations and people have suggested experiments that may look for the presence of torsion \cite{lammerzahl,Shapiro,KTR,HEHL20131775,magueijo,Yuri}. Still, there are more general torsionful theories where torsion does propagate (e.g., Ref.~\cite{Troncoso:1999pk}).

This paper considers a setup where metric and torsion are arbitrary and no attention is set on the theory and fields configuration that can give rise to such geometrical fields. In other words, the results of this paper are purely geometrical. The goal is to generalize a method that is well known in GR to obtain conserved quantities along autoparallels.

It should be stressed that, in most theories in the framework of Riemann--Cartan geometries, classical free test point-like particles follow geodesics, and only particles with intrinsic spin, which are regarded as not point-like, are sensitive to torsion \cite{hehl1971does,trautman1972einstein,Yasskin,Audretsch,Nomura}. Nevertheless, it may be the case that in other torsionful modified theories of gravity some particles do follow autoparallels, or that one is interested in finding the autoparallels to study some geometrical or topological aspects of spacetime. In fact, recently, people are placing attention to the role of torsion in the Raychaudhuri equation \cite{Raychaudhuri1, Raychaudhuri2}, which calls for the use of autoparallel congruences, and on the use of Killing horizons to study torsion effects on thermodynamical quantities \cite{Dey}. Another interesting case where the results of this paper can be applied are the Teleparallel theories of gravity \cite{Aldrovandi} where all the gravitational degrees of freedom are encoded in a torsionful connection.

The paper is organized as follows: in Sec.~\ref{Prel} introduces basic concepts and some preliminary results are discussed. Sec.~\ref{Theorem} is the core of this paper where the definitions are given and the main results are proven. Then, Sec.~\ref{Ex} is devoted to study a particular example. Finally, the concluding remarks are presented in Sec.~\ref{concl}. Even tough many of the results of this paper are valid in arbitrary dimensions and arbitrary signatures, the framework of this work is a Lorentzian geometry in $N$ dimensions where the metric signature is $(-1,+1,\ldots,+1$). The notation and conventions of Ref. \cite{Wald} are used throughout the text, in particular, when convenient, indices are lowered and raised with the metric $g_{ab}$ and its inverse $g^{ab}$.

\section{Preliminaries}\label{Prel}

In this section, some well-known mathematical objects that are used throughout the text are introduced. A connection is a tensor ${C^c}_{ab}$ that links two derivative operators \cite[chapter 3.1]{Wald}. If the operators under consideration are $\mathcal{D}_a$, a torsionful and metric compatible derivative, and $\partial_a$, the partial derivatives associated with some coordinates, then the corresponding connection takes the form
\begin{equation}
{C^c}_{ab}={\Gamma^c}_{ab}+{K^c}_{ab},
\end{equation}
with ${\Gamma^c}_{ab}$ being the conventional Christoffel symbols and ${K^c}_{ab}$ the so-called contorsion tensor. These objects satisfy
\begin{eqnarray}
{\Gamma^c}_{ab}&=& \frac{1}{2}g^{cd}\left(\partial_a g_{db} +\partial_b g_{ad}-\partial_d g_{ab}   \right),\\
{K^c}_{ab}&=&\frac{1}{2}\left ({T^c}_{ab}+g^{cd}g_{be}{T^e}_{da}+g^{cd}g_{ae}{T^e}_{db} \right),
\end{eqnarray}
where ${T^c}_{ab}={T^c}_{[ab]}$ is the torsion tensor and the metric is explicitly written to avoid confusions. Note that ${C^c}_{[ab]}={T^c}_{ab}/2$ and ${K^c}_{(ab)}\neq 0$. Of course, one can also consider the covariant derivative operator $\nabla_a$, which is metric compatible and torsion free; this operator is linked with $\mathcal{D}_a$ by the contorsion.

Geodesics are the curves that extremize the spacetime distance. The geodesic equations are $N$ coupled second-order differential equations, and finding solutions is, in general, very hard. Thus, any method to find solutions to such equations is extremely valuable. Perhaps the most popular method in these regards is the use of symmetries and conserved quantities. In the framework of GR, symmetries are associated with Killing vectors, which give rise to conserved quantities. The method to use these conserved quantities to simplify the geodesic equations solution is described in most GR textbooks, see, e.g., Refs.~\cite{Wald,Hartle}.

In the torsionless case, geodesics and autoparallels, namely, the curves for which the tangent is parallelly transported along them, coincide. However, when a nontrivial torsion is present, these curves differ from each other. The tangent vector $u^a$ of the autoparallels satisfies $0=u^b\mathcal{D}_b u^a $, which can be casted into
\begin{eqnarray}\label{autoparallel}
0&=&\frac{d u^a}{d \lambda}+ {C^a}_{(bc)}u^c u^d\nonumber\\
&=&\frac{d u^a}{d \lambda}+ {\Gamma^a}_{bc}u^c u^d+{K^a}_{(bc)}u^c u^d,
\end{eqnarray}
where $\lambda$ is the corresponding affine parameter. Moreover, the fact that $\mathcal{D}_a$ is metric compatible implies that the causal nature of the curve does not change along it.

On the other hand, the geodesic equations take the form
\begin{equation}\label{geod}
0=u^b\nabla_b u^a=\frac{d u^a}{d \lambda}+ {\Gamma^a}_{bc}u^c u^d=0.
\end{equation}
The fact that these are torsion-independent equations reflects the fact that extremizing spacetime distances is a purely metrical condition. Importantly, Eqs.~\eqref{autoparallel} and \eqref{geod} coincide if and only if
\begin{equation}\label{Delta}
{T^c}_{ba} u^bu_c=0,
\end{equation}
along the curve.

In general, the autoparallel equations are also hard to solve since, after all, they are similar to the geodesic equations. The results of this paper, which are presented in the next section, can be used to significantly reduce such a complicated task when symmetries are present.

\section{T-Killing vectors}\label{Theorem}

A Killing vector $\zeta^a$ is a vector field such that the Lie derivative of the metric along it vanishes, namely, $\nabla_{(a}\zeta_{b)}=0 $. One can then prove the following\\
 \textbf{Theorem (Killing):} If $\zeta^a$ is a Killing vector, then $u^a \zeta_a$ is constant along a geodesic with tangent $u^a$.\\
The proof of this theorem is well known and it is a particular case of the proof given below, hence, it is omitted.

The goal of this work is to generalize the previous theorem in the presence of torsion. A vector field $\xi^a$ is called a T-Killing vector, where the T is for torsion, if
 \begin{equation}
  \mathcal{D}_{(a}\xi_{b)} = 0.
  \label{T-Killing}
 \end{equation}
With this definition it follows\\
 \textbf{Theorem (T-Killing):} If $\xi^a$ is a T-Killing vector, then $u^a \xi_a$ is constant along an autoparallel with tangent $u^a$. \\
 \textbf{Proof.}
 The change of $u^a \xi_a$ along the autoparallel is given by 
 \begin{eqnarray}
  u^a \mathcal{D}_a (u^b \xi_b) &=& (u^a  \mathcal{D}_a u^b) \xi_b + u^a u^b  \mathcal{D}_a \xi_b \nonumber\\
                      &=& u^{a} u^{b} \mathcal{D}_{a} \xi_{b} \nonumber\\
                      &=& 0,
 \end{eqnarray}
where the autoparallel equation and Eq.~\eqref{T-Killing} are used. This completes the proof.

How many independent T-Killing vectors can there be? The analysis presented here closely follows the counting of Killing vectors (see Ref. \cite[Appendix C]{Wald}). The curvature tensor associated with $\mathcal{D}_a $, ${R_{abc}}^d$, is defined by
\begin{equation}
{R_{abc}}^d \omega_d=\left(\mathcal{D}_a\mathcal{D}_b-\mathcal{D}_b\mathcal{D}_a+{T^d}_{ab}\mathcal{D}_d\right)\omega_c ,
\end{equation}
where $\omega_a$ is an arbitrary covector. If one takes $\omega_a=g_{ab} \xi^b$, with $\xi^a$ a T-Killing vector, then
\begin{equation}
{R_{abc}}^d \xi_d = \mathcal{D}_a\mathcal{D}_b\xi_c +\mathcal{D}_b\mathcal{D}_c\xi_a +{T^d}_{ab}\mathcal{D}_d\xi_c ,
 \end{equation}
 where Eq.~\eqref{T-Killing} is used. Interchanging the indices of the last equation and adding up these equations yields
\begin{eqnarray}
\left({R_{abc}}^d +{R_{bca}}^d+{R_{acb}}^d\right)\xi_d &=& 2\mathcal{D}_b\mathcal{D}_c\xi_a +{T^d}_{ab}\mathcal{D}_d\xi_c \nonumber \\&&+{T^d}_{bc}\mathcal{D}_d\xi_a-{T^d}_{ca}\mathcal{D}_d\xi_b,\nonumber\\
 \end{eqnarray}
which can be rewritten as
\begin{eqnarray}
\mathcal{D}_a\mathcal{D}_b\xi_c&=&  -{R_{bca}}^d\xi_d+{T^d}_{bc}\mathcal{D}_d\xi_a+ \frac{3}{2}{R_{[abc]}}^d\xi_d\nonumber\\&&-\frac{3}{2}{T^d}_{[ab}\mathcal{D}_{|d|}\xi_{c]},
 \end{eqnarray}
where, in contrast with the torsion free case, ${R_{[abc]}}^d$ is generically nonzero.

Now, let $L_{ab}\equiv \mathcal{D}_{a} \xi_b=L_{[ab]}$. Then there is a system of equations that, given $\xi_a$ and $L_{ab}$ at a spacetime point, allows one to integrate along a curve with tangent $v^a$. Such a system takes the form
\begin{eqnarray}
\label{dxi}v^a \mathcal{D}_{a} \xi_b&=& v^a L_{ab},\\
\label{dL}v^a \mathcal{D}_{a} L_{bc}&=&v^a\left(-{R_{bca}}^d + \frac{3}{2}{R_{[abc]}}^d\right)\xi_d\nonumber\\
&&+v^a\left({T^d}_{bc}\delta^e_a-\frac{3}{2}{T^d}_{[ab}\delta_{c]}^e\right)L_{de}.
\end{eqnarray}
Therefore, if curvature and torsion are known, then $\xi_a$ and $L_{ab}$ at a point $p$ determine the value of these tensor fields in a neighborhood of $p$. Since, the total number of independent components of $\xi_a$ and $L_{ab}$ at a point is $N$ and $N(N-1)/2$, respectively, which adds up to $N(N+1)/2$, the maximum number of independent T-Killing vectors is $N(N+1)/2$. Moreover, as a corollary, if $\xi_a=0=L_{ab}$ at a point, then $\xi_a=0$ everywhere.
 
A counting of the maximum number of T-Killing vectors has been given. Interestingly, the maximum number coincides with that number for Killing vectors. One could then ask: When is a standard Killing vector a T-Killing vector? Recalling that 
\begin{equation}\label{T vs noT}
\mathcal{D}_{(a}\rho_{b)} = \nabla_{(a}\rho_{b)} - {K^d}_{(ab)}\rho_d,
\end{equation}
it is easy to verify that, if $\rho^a$ is a Killing vector, then it is also a T-Killing vector if and only if ${K^d}_{(ab)}\rho_d = 0 $. Furthermore, it is clear from this result that, when torsion vanishes, all T-Killing vectors reduce to Killing vectors. Note, however, that not all T-Killing vectors must be Killing vectors since, when nonzero, the terms on the right-hand side of Eq.~\eqref{T vs noT} can cancel out. 

What is more, it is known that the commutator of Killing vectors is itself a Killing vector. Interestingly, the commutator of T-Killing vectors is generically not a T-Killing vector. In fact, it is possible to verify that, even if $\rho^a$ and $\sigma^a$ are Killing vectors such that ${K^d}_{(ab)}\rho_d = 0={K^d}_{(ab)}\sigma_d $, generically ${K^d}_{(ab)}[\rho,\sigma]_d \neq 0$.
 
Another straightforward generalization of Killing vectors is that of Killing tensors, which, in cases like Kerr spacetime \cite{Walker}, are extremely useful to solve the geodesic equations. This definition can also be extended to the case where there is nontrivial torsion. A completely symmetric $(0,l)$ tensor $\Xi_{a_1\ldots a_l}$ is called a T-Killing tensor if $ \mathcal{D}_{(b}\Xi_{a_1\ldots a_l)} = 0 $. A trivial extension of the proof given above allows one to verify that $\Xi_{a_1\ldots a_l}u^{a_1}\dots u^{a_l}$ is constant along an autoparallel with tangent $u^a$. T-Killing--Yano tensor can also be defined and it is easy to show, following Ref.~\cite{Batista}, that they lead to conserved quantities. 

The best example of the use of symmetries to solve the geodesic equations is in the Schwarzschild spacetime where such a problem is reduced to an effective one-dimensional problem. In the next section, as an example, the same construction is done for autoparallels in the presence of curvature and torsion.
 
\section{Example: Schwarzschild-like geometry with torsion}\label{Ex}

Loosely speaking, a spacetime is spherically symmetric if every tensor involved in its description has vanishing Lie derivative along the $\mathfrak{so}(3)$ generators. On the other hand, a spacetime is static if there exists a timelike hypersurface-orthogonal Killing field. If these symmetries are assumed, then, in the naturally adapted coordinates $t,r,\theta,\phi$, the metric takes the form \cite[Chapter 6]{Wald}
\begin{equation}
 \text{d}s^2 = -f^2(r) \text{d}t^2 + h^2(r)\text{d}r^2 + r^2(\text{d}\theta^2 +\sin^2\theta \text{d} \phi^2),
 \label{metrica}
\end{equation}
where $f(r)$ and $h(r)$ are arbitrary functions. With the same procedure one can find that, in spherical symmetry, torsion has 8 arbitrary functions of $r$ \cite{Ho}. Moreover, after imposing staticity, only 4 functions are left. Thus, the most general static spherically-symmetric torsion reads
\begin{align*}
 {T^t}_{tr}  &=  F_1(r),                &{T^r}_{\theta\phi}  &=  F_2(r)\sin\theta  , \\
 {T^{\theta}}_{r\theta}  &=  F_3(r),    &{T^{\theta}}_{r\phi}  &=  F_4(r)\sin\theta, \\
 {T^{\phi}}_{r\theta}  &=  -F_4(r)\csc\theta,    &{T^{\phi}}_{r\phi}  &=  F_3(r) .
\end{align*}

The goal of this section is to find the T-Killing vectors in a spacetime that has nontrivial torsion and that is spherically symmetric and static. Since the purpose of this section is to illustrate the utility of the method in a simple example, the case where $F_i=0$, for $i=2,3,4$, is considered. For notational simplicity, the only nonvanishing function, $F_1$, is renamed as $F_1\equiv F$. This choice is particularly convenient since, in this geometry, autoparallels lie on the $\theta=\pi/2$ plane. However, it is not of the type where the dynamics enforces $f^2=h^{-2}$ \cite{Salgado,Jacobson}. Importantly, for a generic static spherically-symmetric torsion, autoparallels do not remain on a plane.

At this stage it is possible to check that the $\mathfrak{so}(3)$ Killing vectors are, automatically, T-Killing vectors. However, the timelike T-Killing vector is proportional to the timelike Killing vector $(\partial/\partial t)^a$, concretely,
\begin{equation}
\xi^a =\mathcal{R}\left(\frac{\partial}{\partial t}\right)^a,
\end{equation}
with $\mathcal{R} = \mathcal{R}(r)$ dimensionless and such that
\begin{equation}\label{Req}
\mathcal{R}' =  \mathcal{R}F ,
\end{equation}
where the prime denotes differentiation with respect to the argument, as usual.

The (nonnegative) conserved quantities associated with the timelike and the remaining $\mathfrak{so}(3)$ T-Killing vectors, $\xi^a$ and  $\psi^a\equiv (\partial/\partial \phi)^a$, can be defined, respectively, as
\begin{eqnarray}
E&=&-\xi^a u_a=f^2 \mathcal{R}\dot{t},\\ 
L&=&\psi^a u_a =r^2\dot{\phi},
\end{eqnarray} 
where the overdot represents the derivative with respect to the affine parameter. Note that $E$ is dimensionless and $L$ has dimensions of $r$. From this point on, attention is set on timelike autoparallels; the method can be easily generalized for other cases. These autoparallels can be parametrized in such a way that $g_{ab}u^au^b=-1$. This parametrization condition can be rewritten, when the conserved quantities are used to replace $\dot{t}$ and $\dot{\phi}$, in the form
\begin{equation}\label{eff eq 1}
-1=-\frac{E^2} {f^2\mathcal{R}^2}+h^2\dot{r}^2+\frac{L^2}{r^2}.
\end{equation}
This is already an effective one-dimensional problem that is more tractable than solving the autoparallel equations in its original form.

Further insights can be obtained in particular cases. As an example, the Schwarzschild metric is assumed, where $f^2 =1 -2M/r=h^{-2}$, $M$ being the Schwarzschild mass, and the torsion function is taken as $F=- M/(r^2f^2)$. This function is chosen such that $\mathcal{R}$ is inversely proportional to $f$, thus making the $E$ term in Eq.~\eqref{eff eq 1} independent of $r$. In fact, it is easy to check that $\mathcal{R} = \alpha f^{-1}$, where $\alpha$ is an integration constant that, from this point on, is absorbed into $E$. With all this, Eq.~\eqref{pot eff} can be brought to the form
\begin{equation}\label{pot eff}
\frac{1}{2}E^2=\frac{1}{2}\dot{r}^2+V_{\rm eff}(r),
\end{equation}
where
\begin{equation}
V_{\rm eff}=\frac{1}{2}\left(1-\frac{2M}{r}\right)\left(1+\frac{L^2}{r^2}\right)+\frac{ME^2}{r}.
\end{equation}
Note that $V_{\rm eff}$ contains the effective potential of the torsion-free Schwarzschild solution, which is comprised by a constant term, the Newtonian gravitational potential, the centrifugal term, and a relativistic correction that involves $M$ and $L$, but it also has a torsion correction: $ME^2/r$.

\begin{figure}[t]
\begin{center}
\includegraphics[width=\columnwidth]{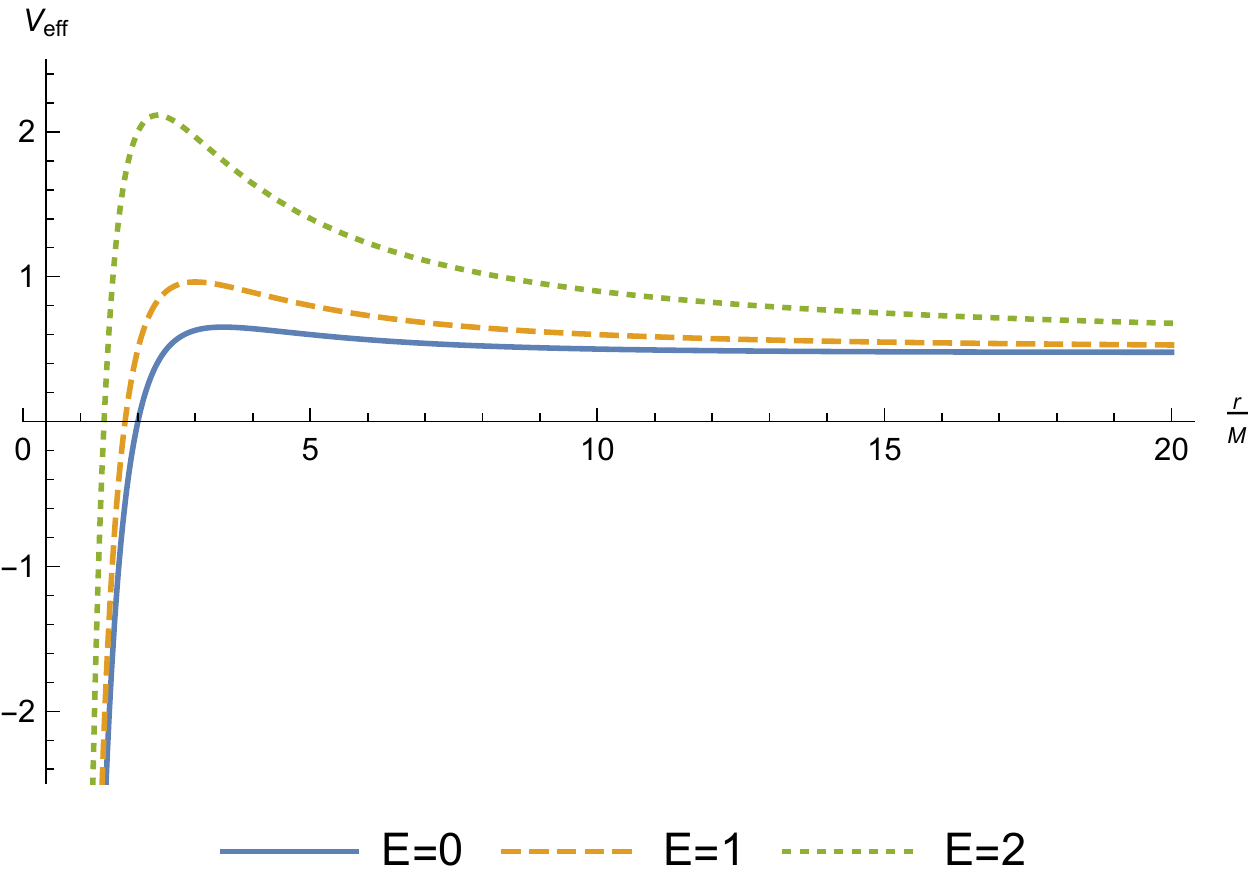}
\end{center}
\caption{\label{fig}(Color online). Effective potential as a function of $r/M$ for $L=5M$ and different values of $E$.}
\end{figure}

The fact that the effective potential depends on $E$, which does not occur in the torsionless case, is a new and interesting feature. Note  that, at the level of the effective potential, the torsion free results are recovered by setting $E=0$. Moreover, the extrema of $V_{\rm eff}$ are located at 
\begin{equation}
r_\pm=\frac{L^2\pm L \sqrt{L^2-12 M^2(1-E^2)}}{2 M\left(1-E^2 \right)}.
\end{equation}
This implies that there are no extrema when $L^2-12 M^2(1-E^2)<0$, and thus, there are no solutions in this regime that are ``trapped'' between to finite radii. Conversely, in contrast to the torsion free case, $L^2-12 M^2(1-E^2)$ can be positive for any $L$. What is more, the height of the maximum of $V_{\rm eff}$ is extremely sensitive to $E$, as can be seen in Fig.~\ref{fig} where $V_{\rm eff}$ is plotted for several values of $E$. 

In this section, a concrete implementation of the ideas developed during the paper is presented. It is clear from the effective potential analysis that torsion can have important effects on the autoparallels, but, technically, extracting information about the autoparalles is not harder than doing it in the torsion free case.

\section{Conclusions}\label{concl}

Symmetries play a key role in modern physics and the solution of the autoparallel equations is no exception. With the definition given here, the problem of solving such equations becomes dramatically simpler in the presence of symmetries. The key concept is that of T-Killing vectors, which allows one to do simple, yet elegant, generalizations of most results associated with Killing vectors.

Of course, in current geometrical theories of gravity geodesics play a more important role than autoparallels. However, it may be the case that some mathematical definitions will be extended to autoparallels, or new techniques will be developed based on these curves, and the present work will play a significant role when symmetries are present. 

\begin{acknowledgments}
We acknowledge having fruitful conversations with Crist\'obal Corral. This research was funded by UNAM-DGAPA-PAPIIT Grant No. IA101818 and by CONACyT through the graduate school scholarships.
 \end{acknowledgments}

\bibliography{References}

\end{document}